Title

# Electrotoroidicity: New Paradigm for Transverse Electromagnetic Responses

Kai Du[1], Daegeun Jo[2], Xianghan Xu[1], Fei-Ting Huang[1], Ming-Hao Lee[3], Ming-Wen Chu[3], Kefeng Wang[1], David Vanderbilt[1], Hyun-Woo Lee[2], and Sang-Wook Cheong[1]*

1. Department of Physics and Astronomy, Rutgers University, Piscataway, New Jersey 08854, USA
2. Department of Physics, Pohang University of Science and Technology, Pohang, Kyungbuk 37673, Republic of Korea
3. Center for Condensed Matter Sciences, National Taiwan University, Taipei 106, Taiwan

* To whom the correspondence should be addressed. (E-mail: sangc@physics.rutgers.edu)

**Abstract**
The exploration of transverse electromagnetic responses in solids with broken spatial-inversion (*I*) and/or time-reversal (*T*) symmetries has unveiled numerous captivating phenomena, including the (anomalous) Hall effect, Faraday rotations, non-reciprocal directional dichroism, and off-diagonal linear magnetoelectricity, all within the framework of magnetotoroidicity. Here, we introduce a novel class of transverse electromagnetic responses originating from electrotoroidicity in ferro-rotational (FR) systems with preserved *I* and *T* symmetries, distinct from magnetotoroidicity. We discover a high-order off-diagonal magnetic susceptibility of FR domains and a reduced linear diagonal magnetic susceptibility at FR domain walls in doped ilmenite $FeTiO_3$. The non-trivial "Hall-like" effect of the former corresponds to an anomalous transverse susceptibility in the presence of spontaneous electrotoroidal moments in FR materials. Our findings unveil an emergent type of transverse electromagnetic responses even in *I* and *T* symmetry-conserved conditions and illustrate new functionalities of abundant FR materials.

**Main text**
Transverse electromagnetic responses are fundamental to electromagnetism as dictated by Maxwell's equations. For example, they explain that a changing electric field *E* can induce a transverse magnetic field *H* (and vice versa) even in vacuum, allowing electromagnetic waves to propagate along the third direction with a velocity

*k*. As for condensed matter systems, transverse electromagnetic responses to external driving fields are often overshadowed by dominant longitudinal responses. However, they can emerge in materials with certain broken symmetries, revealing rich physics beyond the conventional diagonal responses. For example, when time-reversal (*T*) symmetry is broken, various famed transverse responses can occur, such as the Hall effect[1] in magnetic field (*H*) or the anomalous Hall effect[2] with spontaneous magnetizations (*M*), and Faraday rotations in ferromagnets[3]. Recently, it was theoretically predicted[4] and experimentally verified[5] that transverse effects such as the nonlinear Hall effect can occur even without the *T* symmetry breaking if the spatial-inversion (*I*) symmetry is broken instead. When both *I* and *T* symmetry are broken, systems with magnetic toroidal order can also exhibit novel transverse effects such as off-diagonal linear magnetoelectricity[6] and non-reciprocal directional dichroism even for unpolarized light[7]. These transverse effects of transport properties, optical rotations, and magnetoelectricity have been intensely studied in the past as the foundation of each relevant field, continually pushing the frontier of condensed matter physics[8–13].

Intriguingly, all of these known transverse electromagnetic effects under conditions of broken *I* or *T* symmetries (or both) are deeply connected to the orthogonal configuration among the electric dipole (*P*, or electric field *E*), magnetic dipole (*M*, or magnetic field *H*), and velocity vector *k* (or magnetic toroidal moment $T_o$[14]) (Fig. 1a). Specifically, when free spatial rotations are allowed, the orthogonal configuration of these three objects is invariant under all other symmetry operations, such as *I* and *T* (Supplementary Information Fig. S1a). We define the universal invariance of this configuration as "magnetotoroidic kinetics" or simply "magnetotoroidicity". Then, one can demonstrate that all existing transverse electromagnetic responses can be well-explained by this magnetotoroidic invariance, representing a singular type of effects with broken *I* or *T* symmetries (or both) (see Supplementary Information Note 1). Despite these established transverse electromagnetic effects relevant to magnetotoroidicity, a fundamental question remains outstanding: Are new types of transverse electromagnetic effects, stemming from other origins with preserved *I* and *T* symmetries, also possible in condensed matter?

Ferro-rotational (FR) order[15–20] (also known as ferro-axial order or 2D chirality) is an emerging ferroic phenomenon characterized by a spontaneous toroidal arrangement of electric dipoles, whose order parameter is an axial-type electrotoroidal moment *A* with unbroken *I* and *T* symmetries (Fig. 1b). Encouragingly, the electrotoroidal moment *A* of FR order does form a similar robust symmetry-invariant orthogonal configuration when combined with the perpendicular magnetic field and transverse magnetization, which can also harmonize with perpendicular electric field/velocity and transverse polarization/velocity, respectively (Fig. 1b). Thus, one can also define these symmetry-invariant configurations (Fig. S1b-1d) as "electrotoroidicity", similar to magnetotoroidicity. Emphasize that the critical distinction here lies in the unbroken *I* and *T* symmetries of electrotoroidal moment *A*, enabling a previously undiscovered category of transverse electromagnetic responses. A signature effect of the electrotoroidicity can be the never-observed anomalous transverse susceptibility (ATS) based on the electrotoroidal moment $A$ in FR materials (Fig. 1b). Therefore, FR systems are ideal platforms for exploring this new class of transverse electromagnetic responses originating from electrotoroidicity with unbroken $I$ and $T$ symmetries. With the recent advances in research on FR materials, the potential existence of ATS may also introduce susceptibility as a new fundamental physical property capable of exhibiting non-trivial transverse responses, alongside those known transverse transport properties, optical rotations, and transverse linear magnetoelectricity. Apparently, ATS mirrors the odd-order characteristics of the Hall effect, where switching the direction of driving field or the electrotoroidal moment $A$ will change the sign of the transverse response [21] (Fig. S2a). We will demonstrate later that it can be a high-odd-order transverse effect depending on the specific symmetry of the material. A detailed analysis of the symmetry requirements for such non-trivial odd-order ATS can be found in Fig. S2b. Importantly, we elaborate the distinct difference between the trivial birefringence-type linear off-diagonal response and non-trivial "Hall-like" response, such as the standard (anomalous) Hall effect and the ATS, in Supplementary Information Note 2. Hereafter, we focus on the magnetic version of non-trivial ATS, which is closely tied to the electrotoroidal moment of FR order (Fig. 1b).

Experimentally, transverse magnetic responses are rarely observed. Only a possible second-order off-diagonal magnetic susceptibility was proposed in magnetoelectric $LiNiPO_4$ so far[22], while its mechanism remains completely unclear. Recently, the concept of ATS has been theoretically proposed[21,23]. However, the experimental verification of this "Hall-like" odd-order transverse susceptibility has yet to be achieved. The main obstacle comes from the difficulty in characterizing FR materials, partially due to the lack of inversion symmetry breaking and the complexity of FR domains[17,18,20]. In this letter, we overcome these challenges with comprehensive studies on a magnetic FR material $Fe_{1.23}Ti_{0.77}O_3$ and report the observation of its high-odd-order ATS at room temperature as a new class of transverse electromagnetic response originating from electrotoroidicity.

**Results**

**Ferro-rotational order and magnetism of Fe-doped iron ilmenite**
$Fe_{1.23}Ti_{0.77}O_3$ is a Fe-doped iron ilmenite with the same FR space group $R\bar{3}$ [24] to its parent compound $FeTiO_3$. Pure $FeTiO_3$ is an antiferromagnet with an ordered stacking of $Fe^{2+}$ and $Ti^{4+}$ layers, which introduces an imbalance in oxygen rotation and gives rise to a net rotation in each unit cell. Upon partial Fe doping into Ti sites, Fe-doped iron ilmenites become ferrimagnetic[25] while still preserving the FR nature. Its ferrimagnetic $Tc$ rises gradually while the concentration of Fe doping increases[26]. The ability to tune the ferrimagnetic $Tc$ all the way above room temperature makes Fe-doped iron ilmenites an ideal FR system to investigate ATS at room temperature.

Therefore, $Fe_{1.23}Ti_{0.77}O_3$ with a *Tc* near 295 K is chosen in this work to conveniently study its magnetic transition and observe ATS by utilizing its large magnetic susceptibility at room temperature. Depending on the stacking sequence, both clockwise (CW) and counter-clockwise (CCW) ferro-rotational domains can potentially coexist with the corresponding FR axial vector ***A*** parallel/antiparallel to the crystallographic *c* axis (Fig. 1c). Our transmission electron microscopy (TEM) study shows clear coexistence of both CW and CCW FR domains in our high-quality single crystals (Fig. 1d), which will be discussed in the latter section that it is crucial for the observation of potentially ATS. Intriguingly, its Ising-type FR domain size can be controlled by different cooling rates across its FR transition (Fig. 1d), which is consistent with the Kibble–Zurek mechanism (KZM) discovered recently for similar Ising-type FR domains in $NiTiO_3$[19].

As the FR order is recognized as a new ferroic order in addition to conventional ferromagnetic, ferroelectric, ferroelastic, and ferro-toroidal order, its impact on magnetism remains unclear at this point. The coexistence of multiple ferroic orders in single–phase materials, termed "multiferroics"[27], can exhibit richer physics of interplays between different ferroic orders. Therefore, careful characterizations of magnetic properties and potential interplays between the FR domains and ferrimagnetic domains in $Fe_{1.23}Ti_{0.77}O_3$ are desirable before testing its ATS. Our temperature-dependent magnetization curve (Fig. 2a) shows an expected paramagnetic(PM)-to-ferrimagnetic(FM) transition around 295 K with a clear easy-plane anisotropy. At room temperature around 295 K, it is in a superparamagnetic state (sPM) with a huge magnetic susceptibility but no spontaneous long-range orderings. This is also evident from the magnetization curve as a function of magnetic fields (Fig. 2b), where no hysteresis but only a steep magnetization slope is observed at this superparamagnetic state. Strikingly, our magnetic force microscopy (MFM) reveals clear FR domain wall contrasts at room temperature on both the *ab* plane (Fig. 2c) and side surface (Fig. 2d) of $Fe_{1.23}Ti_{0.77}O_3$ when the magnetic field is applied perpendicular to the scanned surface. Based on the contrast polarity, this suggests a significantly reduced diagonal magnetic susceptibility $\chi_{xx}$ of FR domain walls compared to FR domains (Fig. 2e). The conclusion has been further confirmed by consistent FR domain wall signals obtained from both standard soft magnetic tips and non-switchable hard magnetic tips in opposite magnetic fields (Fig. S3). Note that FR domains are generally believed to be difficult to probe due to the non-broken inversion and time-reversal symmetry of FR order. Only limited methods such as TEM[21], linear electrogyration microscopy[18], second-harmonic gerneration[28], and selective polishing[20] are shown to be able to visualize FR domains so far. This reduced $\chi_{xx}$ of FR domain walls not only provides us an easy and non-destructive way to visualize FR domains in $Fe_{1.23}Ti_{0.77}O_3$ but also adds another accessible tool for imaging hard-to-seen FR domains in potentially all magnetic FR materials. Interestingly, our atomic imaging investigation shows that this reduced susceptibility at FR domain walls likely stems from the tendency of having Fe deficiencies at FR domain walls (Fig. S4). Another notable feature is that these FR domains in $Fe_{1.23}Ti_{0.77}O_3$ have an obvious elongation along *ab* plane and a compression along the *c* axis. On the other hand, the domain shape is almost isotropic on the *ab* plane (Fig. 2f). More quantified comparisons of the FR domain shape anisotropy can be found in Fig. S5. Future studies on other FR materials are desired to check if this domain shape anisotropy is a universal hallmark of FR orders.

With the knowledge of FR domains in $Fe_{1.23}Ti_{0.77}O_3$, a comprehensive temperature-dependent MFM study is performed down to the low temperature at zero field (Fig. S6). At temperatures well below $T_c$, stripe-like ferrimagnetic domains with their walls along the hexagonal *ab* plane direction are observed. Unprecedentedly, these ferrimagnetic domains tend to terminate abruptly at certain curved boundaries, which turn out to be FR domain walls. Recently, chiral domain walls that are structurally distinct from the domains they separate have been demonstrated to have a profound effect on magnetic textures[29]. However, how FR domain walls will affect magnetic properties is still elusive so far. Our results indicate a strong impact of FR domain walls on magnetic orders, similar to the impact of chiral domain walls on helical magnetic domains[29]. Upon approaching the superparamagnetic state near room temperature, these ferrimagnetic domains fade away while only faint FR domain wall contrasts remain visible due to the small stray fields from the MFM tip and the aforementioned reduced $\chi_{xx}$ at FR domain walls. Similar MFM images and abrupt termination of ferrimagnetic domains at FR domain walls are observed in the next thermal cycle cooled from above $T_c$, which demonstrates these behaviors as an intrinsic effect. These results exemplify a strong coupling between the FR and ferrimagnetic order and possible exotic domain wall magnetism in magnetic FR materials.

**High-order anomalous transverse susceptibility (ATS) and thermodynamics analysis**

Having established the FR nature and conventional linear diagonal magnetic response of $Fe_{1.23}Ti_{0.77}O_3$, we now examine its off-diagonal ATS at room temperature where its magnetic response is supposedly maximized. As off-diagonal Hall-like signals are often much smaller than the dominant diagonal component, it is critical to eliminate inevitable contaminations from longitudinal signals. Typically, Hall measurements adapt an antisymmetrization process by sweeping a full magnetic field loop to subtract the longitudinal component. However, such a subtraction protocol is not available for the ATS since the conjugate field of FR order is still lacking so far [17], and getting a full field loop of ATS is impractical. Fortunately, one can directly probe the ATS utilizing spontaneous FR domains even in the absence of the conjugate field of FR order. As illustrated in Fig. S2a, the aligned diagonal magnetic response of CW and CCW FR domains parallel to the external magnetic field ***H*** will not contribute to the magnetic contrast while the opposite off-diagonal magnetic response can be exclusively picked up by MFM. Fig. 3a shows a typical MFM image taken with a high-coercivity tip[30] on the side surface of $Fe_{1.23}Ti_{0.77}O_3$ at 295 K in the presence of an in-plane magnetic field ($***H_x***$= +1 kOe) perpendicular to FR axial vector ***A***. Clear out-of-plane domain contrasts are observed, which are consistent with the FR domain pattern in the circular differential interference contrast (cDIC) image revealed by the rotational selective polishing process[19,20] (Fig. 3b). To further confirm these magnetic contrasts are fully in line with ATS of FR order, MFM images with different magnetic field orientations are carefully compared in Fig. 3c-3f. Essentially, no obvious magnetic signals are observed in the zero field condition (Fig. 3c) or when the field is parallel to the FR axial vector *A* (Fig. 3e), which can safely exclude extrinsic origins of observed magnetic signals. Only faint domain wall signals are barely visible due to the aforementioned reduced diagonal susceptibility at FR domain walls. Most importantly, expected domain contrast reversal is evident when magnetic fields are applied perpendicular to the FR axial vector *A* in opposite directions (Fig. 3d and 3f). These observed off-diagonal magnetic responses of $Fe_{1.23}Ti_{0.77}O_3$ are fully in accord with the odd-order ATS of FR order proposed in Fig. S2a.

It is worth noting that FR order in general allows all odd-order ATS including high-order ones. Although the contrast reversal could confirm the observed ATS in $Fe_{1.23}Ti_{0.77}O_3$ as an odd-order effect, additional analysis is needed to determine its exact order[23]. Specifically, the thermodynamic relation

$$M_i = -\frac{\partial f}{\partial H_i}, \tag{1}$$

imposes severe constraints on the order, where *f* is the free energy density. For the linear ATS, one can have the susceptibility tensor $\chi_{ij}$ defined by

$$\begin{pmatrix} M_x \\ M_y \end{pmatrix} = \begin{pmatrix} \chi_{xx} & \chi_{xy} \\ \chi_{yx} & \chi_{yy} \end{pmatrix} \begin{pmatrix} H_x \\ H_y \end{pmatrix}. \tag{2}$$

Then, Eq. (1) requires off-diagonal components of $\chi_{ij}$ to be symmetric since

$$\chi_{xy} = \left(\frac{\partial M_x}{\partial H_y}\right)_{H=0} = -\left(\frac{\partial^2 f}{\partial H_y \partial H_x}\right)_{H=0} = -\left(\frac{\partial^2 f}{\partial H_x \partial H_y}\right)_{H=0} = \left(\frac{\partial M_y}{\partial H_x}\right)_{H=0} = \chi_{yx}. \tag{3}$$

In materials with (approximate) continuous rotation symmetry, on the other hand, the symmetry implies $\chi_{xy} = -\chi_{yx}$, which contradicts the implication of Eq. (3), forbidding the linear order ATS. The same conclusion holds when the continuous symmetry is replaced by $C_{3z}$ or $C_{4z}$ (Supplementary Information Note 3). Since ideal $Fe_{1.23}Ti_{0.77}O_3$ has $C_{3z}$, the observed off-diagonal magnetic response of $Fe_{1.23}Ti_{0.77}O_3$ should not be linear. Our analysis (see Supplementary Text for details) indicates that the thermodynamic relation makes the ATS fifth-order response in materials with $C_{3z}$. This is vastly different from the anomalous Hall effect described by the conductivity tensor σ where *J=σE*, as there is no such thermodynamic relation between current density *J* and electric field *E* like Eq. (1). Therefore, an off-diagonal response $\sigma_{xy}=-\sigma_{yx}$ is generally allowed in the linear regime while the non-trivial $\chi_{xy}$ responsible for linear ATS is forbidden in systems with $C_{3z}$. One exotic feature of the high-order transverse magnetic response in FR materials with $C_{3z}$ is that the rotation direction of the response with respect to the driving field will alternate when the field rotates within the *xy* plane. The complete "transverse response wheels" for $C_{2z}$, $C_{3z}$, and $C_{4z}$ systems with and without FR order are compared in Fig. 4a. For $C_{2z}$ systems, trivial birefringence-type linear transverse susceptibility can exist readily. Additional FR order on top of the $C_{2z}$ will add additional linear contributions and result in the tilting of aforementioned principal axes toward opposite directions for different FR domains. On the other hand, $C_{3z}$ and $C_{4z}$ systems with symmetric transverse susceptibilities do not allow the linear transverse susceptibility even in the presence of FR order, and their leading order of transverse susceptibility are fifth- and third-order, respectively. These interesting facts illustrate that the high-order transverse electric/magnetic response is a unique characteristic of $C_{3z}$ and $C_{4z}$ systems with FR order.

Although the response of a system to the external driving field is often dominated by the first-order linear term, systems with only high-order response can provide exotic and rich physics beyond the conventional linear regime. While the nonlinear optical effect of materials like second-harmonic generations (SHG)[31] and high-harmonic generations (HHG)[32] has been widely studied ever since the introduction of modern strong light sources like high-power lasers, direct observations of high-order electrical or magnetic effects have been very limited so far. The recent discovery of nonlinear Hall effect[4,5] in $WTe_2$ is one prototype example of a second-order electrical effect with vanishing linear response. As of nonlinear magnetic effects, the anomalies of third-order magnetic susceptibility are widely considered as signatures of exotic magnetic states such as quantum spin liquids[33]. As far as we know, a magnetic response up to the fifth order, especially a non-trivial off-diagonal one, has never been experimentally reported before.

**First-principle calculations: Linear response with strain**
The leading order of response may be altered by modifying symmetry characteristics. For instance, the second-order anomalous Hall effect changes to the linear-order anomalous Hall effect by breaking the time-reversal symmetry[4]. Similarly, the leading order of response for the ATS may be altered by modifying symmetry characteristics. Here, we demonstrate theoretically that the leading order of the ATS can be considerably lowered from the fifth-order to the linear order simply by stretching $Fe_{1.23}Ti_{0.77}O_3$ along an in-plane direction, say along the x direction, thereby breaking $C_{3z}$ in the ideal $Fe_{1.23}Ti_{0.77}O_3$. We perform first-principles calculations to examine the $C_{3z}$ breaking effect. For the theoretical study, pure $FeTiO_3$ should be already sufficient to capture FR nature and all the criteria for ATS since the Fe-doping and high $Tc$ of $Fe_{1.23}Ti_{0.77}O_3$ is only beneficial for the experimental observation of ATS. Therefore, the parent compound $FeTiO_3$ was used for calculations for the sake of simplicity without losing the generality (see Methods). The green curve in Fig. 4b shows the calculated off-diagonal magnetic susceptibility $\chi_{xy}$ of CCW FR domain in $FeTiO_3$ in the linear response region with almost vanishing signals. This is consistent with the previous conclusion that linear $\chi_{xy}$ is forbidden by the $C_{3z}$ symmetry. On the other hand, it is worth noting that doped iron ilmenites are reported to have large magnetostriction[34] in the presence of magnetic fields, which may help break such a restriction from $C_{3z}$ symmetry and enable a linear off-diagonal response. However, it will be challenging to treat the strain induced by the magnetic field at the first-principle level. Alternatively, we calculated the linear $\chi_{xy}$ of different FR domains with a 2% strain applied along [110] direction parallel to the magnetic field to mimic magnetostriction (Fig. 4). Considering magnetostriction is a field-even effect, the calculation of such linear magnetic response in the presence of strains along the field direction is effectively mimicking a high-odd-order effect. In this case, non-zero $\chi_{xy}$ can be obtained from the calculation on both FR domains (red curves in Fig. 4b and 4c). Interestingly, the most pronounced effect occurs when the system is slightly hole-doped with $E_F=-0.2$ eV, which matches well with the carrier type[35] and the doping level of experimentally studied $Fe_{1.23}Ti_{0.77}O_3$. Moreover, the sign reversal of calculated $\chi_{xy}$ on opposite FR domains is again fully consistent with our proposed odd-order ATS enabled by FR orders in Fig. 1e. We would also like to point out that the non-FR $R\bar{3}c$ structure does not give rise to such non-trivial transverse susceptibility $\chi_{xy}$ in our calculation even if a similar strain is applied (green curve in Fig. S7). Therefore, $C_{3z}$ symmetry breaking by the magnetostrictive effect itself is not a sufficient condition for the non-zero transverse susceptibility, which again emphasizes the critical role of the electrotoroidicity for ATS.

**Discussion**

Our work has far-reaching implications, suggesting that other fundamental properties—such as transport responses of electrons or phonons—can also exhibit similar transverse effects. They correspond to the off-diagonal $\boldsymbol{k}$ response of the applied $\boldsymbol{k}$ in FR systems (Fig. 1b). Specifically, this transverse transport effect should resemble the standard anomalous/thermal Hall effects. However, the response is characterized by transverse electric/thermal currents, rather than transverse voltage/thermal gradient in (anomalous) Hall/thermal Hall effects. In addition, a similar anomalous "dielectric" transverse susceptibility may also exist in FR materials, where an odd-order transverse electric susceptibility is allowed in response to a longitudinal electric field (Fig. 1b). It corresponds to a "dielectric" version of the ATS reported in this work. While chiral materials have found widespread use in various research disciplines and technologies, potential applications of FR materials, as the 2D counterpart of chiral materials, remain largely unexplored. Our findings introduce a new dimension of fundamental transverse electromagnetic responses in solid-state matter and unveil exciting untapped functionalities of FR materials. For instance, a magnetic field can exert magnetic torques on single-domain FR materials even in their paramagnetic state at or above room temperature, enabling the potential rotation and manipulation of FR nanoparticles (Fig. S8a). By leveraging the same magnetic torque, one could construct FR-material-based motors operating in an AC magnetic field at an appropriate frequency, directly converting magnetic field energy into mechanical energy without any contact frictions. Moreover, since both electric and magnetic susceptibilities exhibit ATS, the same transverse effect should apply to the electric and magnetic fields of light at optical frequencies. Consequently, ferro-rotational systems could exhibit high-order optical rotations of linearly-polarized light in both transmission (Faraday-type) and reflection (Kerr-type) geometries (Fig. S8b). This represents a novel high-order transverse optical response, distinct from the conventional natural optical rotation observed in chiral materials, and could become significant for high harmonic generation[32] applications when light is intense.

In summary, our proof-of-concept results and theoretical calculations established the existence of non-trivial ATS as a novel transverse electromagnetic effect of the FR order with electrotoroidicity, in contrast to all previous transverse effects with broken *I* or *T* symmetries (or both). The emergence of high-order ATS opens an era of exploring a new class of transverse electromagnetic effects under symmetry-preserved conditions and their potential room-temperature applications in abundant FR systems.

## Methods

### Crystal growth
Single crystals of $Fe_{1.23}Ti_{0.77}O_3$ were grown by a floating zone technique. High-purity powders of $Fe_2O_3$, Fe, and $TiO_2$ in molar ratio 0.487 : 0.257 : 0.77 were mixed and sintered at 1200°C for 10 hours in $N_2$ flow with one intermediate grinding. The product was shaped into a rod and sintered at 1230°C for 5 hours in $N_2$ flow. The growth was performed at a rate of 4 mm per hour in $N_2$ flow. The purity and crystallinity of the final product are confirmed by powder X-ray diffraction and Laue diffraction. The specimen for anomalous magnetic Hall effect and MFM measurements were post-annealed from 1200°C to 1000°C at a cooling rate of 200°C per hour in an evacuated quartz tube to obtain a desired FR domain density.

### TEM and selective polishing
The dark-field (DF) TEM imaging was performed on JEOL 2010F electron microscope. The high-angle annular dark-field scanning transmission electron microscopy (HAADF STEM) imaging was conducted on spherical-aberration corrected JEOL 2100FX, with the electron-probe size of 0.9~1 Å. The HAADF images were raw data. All electron microscopes were operated at 200 keV and we used conventional mechanical polishing and Ar-ion milling to prepare the $Fe_{1.23}Ti_{0.77}O_3$ specimens.

### Magnetic measurements and MFM
Magnetic properties of $Fe_{1.23}Ti_{0.77}O_3$ crystals were measured in a Magnetic Property Measurement System (MPMS, Quantum Design). Well-polished *ab*-plane and side surfaces of $Fe_{1.23}Ti_{0.77}O_3$ crystals were scanned using a temperature-variable AFM system (Attocube) in a dual pass mode (lift height ~ 35 nm) with either commercial Co/Cr-coated tips or homemade high-coercivity Co/Pt-coated tips. To supply magnetic fields to the MFM specimen, either a large in-plane or out-of-plane magnetized permanent magnet was placed beneath the specimen. The magnetic field strength was tuned by varying the thickness of the magnet and measured using a Gauss meter before the MFM experiment.

### First-principle calculations
The electronic structure of paramagnetic $FeTiO_3$ was calculated by using density functional theory with the full-potential linearized augmented plane-wave program FLEUR[36,37]. The lattice parameters were chosen as *a*=5.17 Å and *c*=14.1 Å for FR phase with space group $R\bar{3}$, and $a$=5.19 Å and $c$=13.9 Å for non-FR phase with space group $R3c$, respectively[38,39]. In addition to these structures, we considered hypothetical structures, where a strain was incorporated by increasing the $x$-components of the lattice vectors by 2%, to mimic the magnetostriction that breaks $C_{3z}$ symmetry. As an exchange-correlation functional, Perdew-Burke-Ernzerhof functional[40] within the generalized gradient approximation is used. The self-consistent calculations were carried out over the Brillouin zone with an 8×8×8 Monkhorst-Pack **k**-mesh. The spin-orbit coupling was treated within the second variation scheme[41]. Employing the DFT+$U$ method[42] with Coulomb energy parameter $U$=5.0 eV for Fe 3$d$ orbitals, we obtained a band gap of around 2.6 eV, which is consistent with the experimental value[43]. Using the Wannier90 code[44,45], we constructed the maximally localized Wannier functions of Fe 3$d$, Ti 3$d$, and O 2$p$ orbitals for each system, which describes the occupied valence bands as well as the conduction bands. Based on the Hamiltonian, spin, and orbital angular momentum operators established in the Wannier gauge, the Van Vleck contribution to the off-diagonal magnetic susceptibility $\chi_{yx}$ was calculated by the linear response theory:

$$\chi_{yx} = \frac{\mu_0 \mu_B^2}{\hbar^2} \sum_{n,m} \int \frac{d^3\mathbf{k}}{(2\pi)^3} (f_{m\mathbf{k}} - f_{n\mathbf{k}}) \frac{\langle u_{n\mathbf{k}}|(L_y + 2S_y)|u_{m\mathbf{k}}\rangle \langle u_{m\mathbf{k}}|(L_x + 2S_x)|u_{n\mathbf{k}}\rangle}{E_{n\mathbf{k}} - E_{m\mathbf{k}} + i\Gamma},$$

(4)

where $\mu_0$ is the permeability of free space, $\mu_B$ is the Bohr magneton, $|u_{n\mathbf{k}}\rangle$ is the periodic part of the Bloch eigenstate for the band $n$ and crystal momentum **k** with the energy eigenvalue $E_{n\mathbf{k}}$, $f_{n\mathbf{k}}$ is the Fermi-Dirac distribution function for the temperature of 300 K, and $L_\alpha$ and $S_\alpha$ are the $\alpha$-components of the orbital and spin angular momentum operators, respectively. The integration was performed over a 100×100×100 **k**-mesh and the broadening energy $\Gamma$ was assumed to be 25 meV.

## Data availability
The data that support the findings of this study are available from the corresponding author upon reasonable request.

## Acknowledgements
The work at Rutgers was supported by the W. M. Keck Foundation grant to the Keck Center for Quantum Magnetism at Rutgers University and NSF Grant DMR-1954856 (D.V.). The work at Pohang University of Science and Technology was supported by the Samsung Science and Technology Foundation (BA-1501-51).

## Competing interests
The authors declare no competing interests.

## Contributions
S.C. and K.D. initiated and guided the project; X.X. and K.D. prepared the samples; K.W. measured bulk magnetic properties; F.T.H., M.H.L, and M.W.C performed TEM investigation; K.D. did MFM measurements; D.J., and H.W.L. did first-principle calculations. H.W.L. and D.V. contributed to the symmetry analysis; K.D. and S.C. analysed the data and wrote the paper with inputs from all authors.

**Supplementary Information**

Supplementary Information is linked to the online version of the paper at www.nature.com/nature.

**Figure legends**

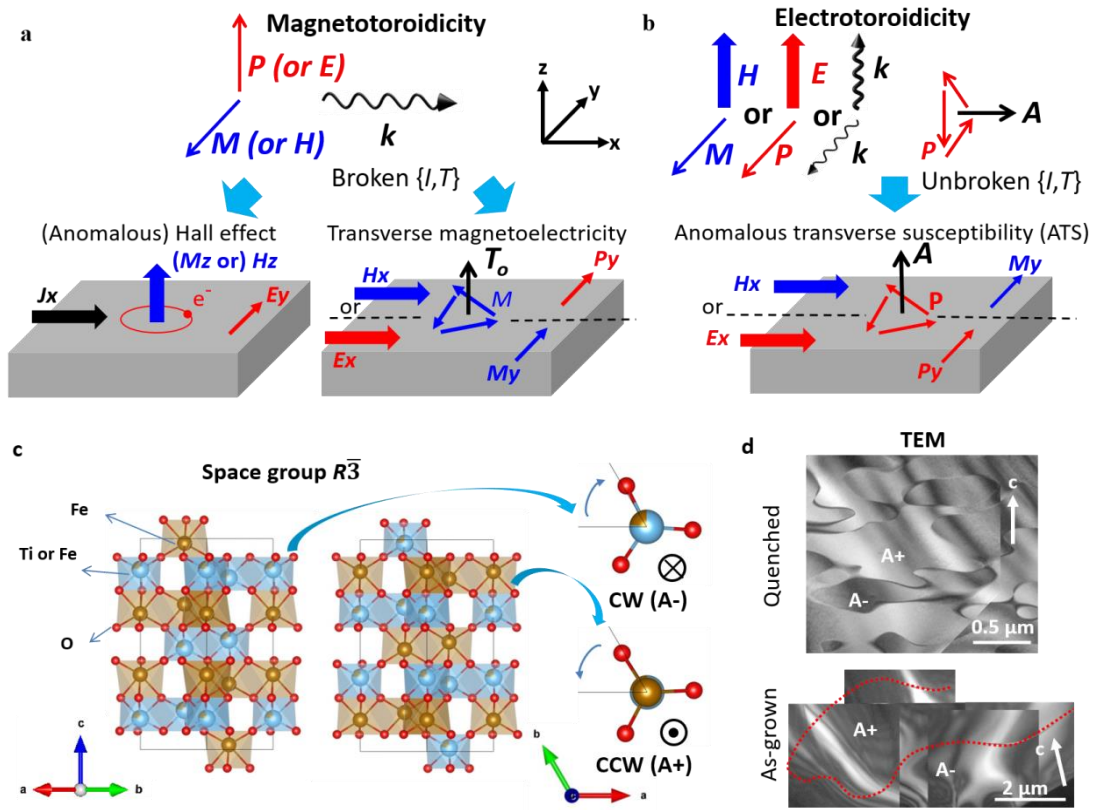

**Figure 1: Schematic of two types of transverse electromagnetic responses. (a)** Schematic of transverse electromagnetic responses originating from "magnetotoroidicity" with broken *I* and *T*, and sellected prototype effects such as the (anomalous) Hall Effect in the presence of (spontaneous magnetizations or) magnetic fields and transverse linear magnetoelectricity with magnetic toroidal order $T_o$. **(b)** Schematic of transverse electromagnetic responses originating from "eletrotoroidicity" with unbroken *I* and *T*, and its prototype effects of novel anomalous transverse susceptibility(ATS) in the presence of spontaneous electrotoroidal moment with the axial-type vector *A*. **(c)** Crystal structure and the two ferro-rotational configurations of $Fe_{1.23}Ti_{0.77}O_3$ with opposite eletrotoroidal moments, A+ and A-. **(d)** Side-view dark-field (DF)-TEM images by selecting spots (113) of air-quenched and as-grown $Fe_{1.23}Ti_{0.77}O_3$ single crystal along $[1\bar{1}0]$. It reveals the coexistence of two types of FR domains, A+ and A-. Those spots are indexed with respect to FR domain A+ and become $(11\bar{3})$ in the reciprocal space with respect to domain A- by the twofold rotation as two FR domains do in real space.

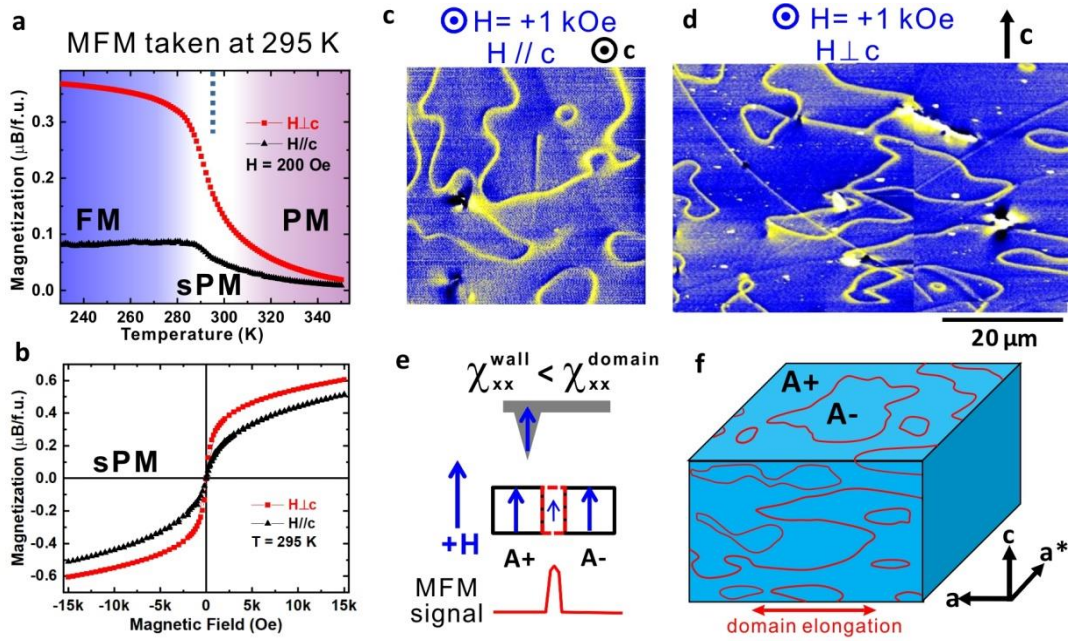

**Figure 2: Magnetic properties and reduced diagonal susceptibility at FR domain walls of $Fe_{1.23}Ti_{0.77}O_3$.** **(a)** Magnetic susceptibility as a function of temperature in 200 Oe with $H\perp c$ and $H//c$, showing the transition from high-temperature paramagnetism (PM) to room-temperature superparamagnetism (sPM), and then to ferrimagnetism (FM) at low temperatures. **(b)** Out-of-plane ($H//$c, black triangles) and in-plane ($H\perp$c, red squares) magnetization curves as a function of the magnetic field in the sPM state at 295 K. MFM image of the hexagonal **(c)** *ab* plane and **(d)** side surface under 1 kOe out-of-plane field at 295 K, showing obvious FR domain wall signals. **(e)** Schematics of reduced diagonal susceptibility at FR domain walls and corresponding MFM signals. **(f)** Schematics of isotropic FR domain pattern on the *ab* plane in comparison to laterally elongated FR domain pattern on the side surface.

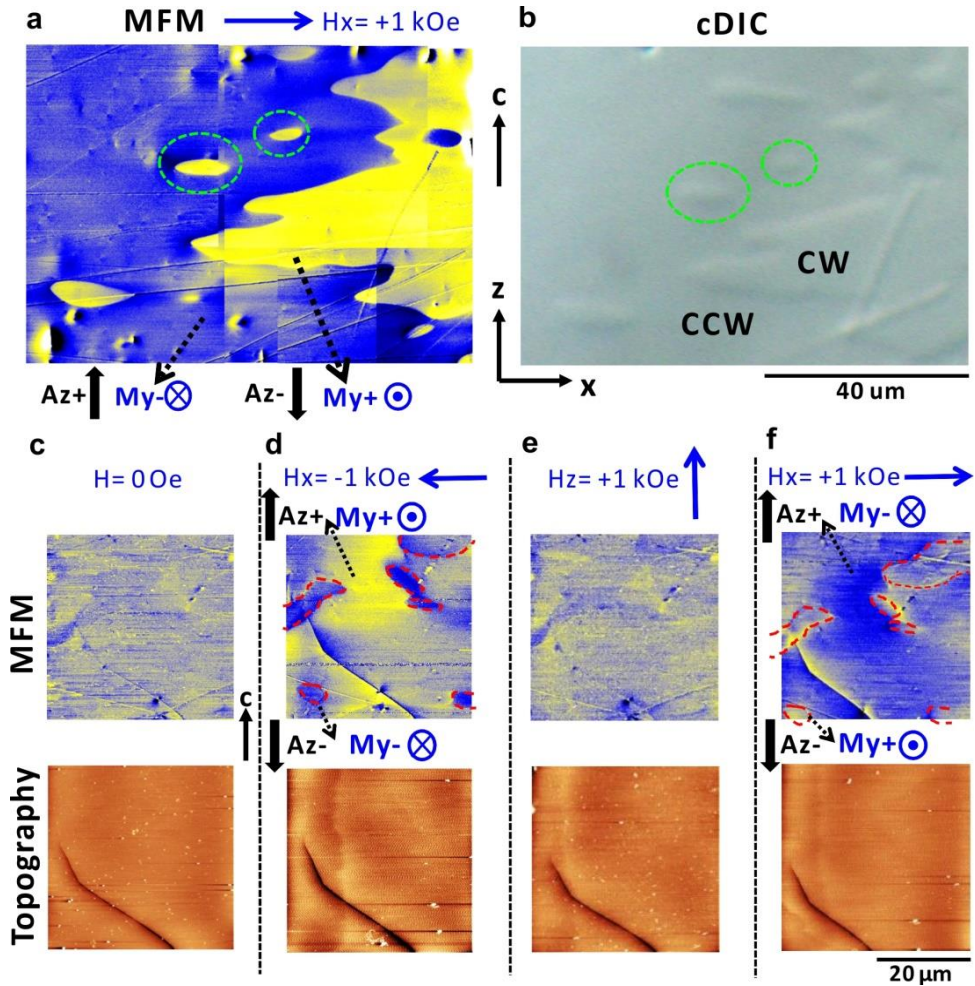

**Figure 3: Observation of odd-order anomalous transverse susceptibility. (a)** MFM image of the side surface of a 200°C/h-cooled $Fe_{1.23}Ti_{0.77}O_3$ with an in-plane field $H_x$=+1 kOe at 295 K, showing anomalous transverse susceptibility with FR domain contrasts. **(b)** cDIC microscope image of the corresponding area in (a) after selective circular polishing with a consistent FR domain pattern. Green dashed circles mark the same FR domains in (a) and (b). MFM and corresponding topography image of the side surface of another 200°C/h-cooled $Fe_{1.23}Ti_{0.77}O_3$ with **(c)** $H$=0 Oe, **(d)** an in-plane field $H_x$=-1 kOe, **(e)** a vertical field $H_z$=+1 kOe parallel to $A_z$, **(f)** an in-plane field $H_x$=+1 kOe, showing contrast reversals consistent with the odd-order anomalous transverse susceptibility. High-coercivity Co/Pt tips are used to ensure the collection of out-of-plane magnetic signals from FR domains in the presence of in-plane fields. Scale bars are 40 μm in (a-b) and 20 μm in (c-f).

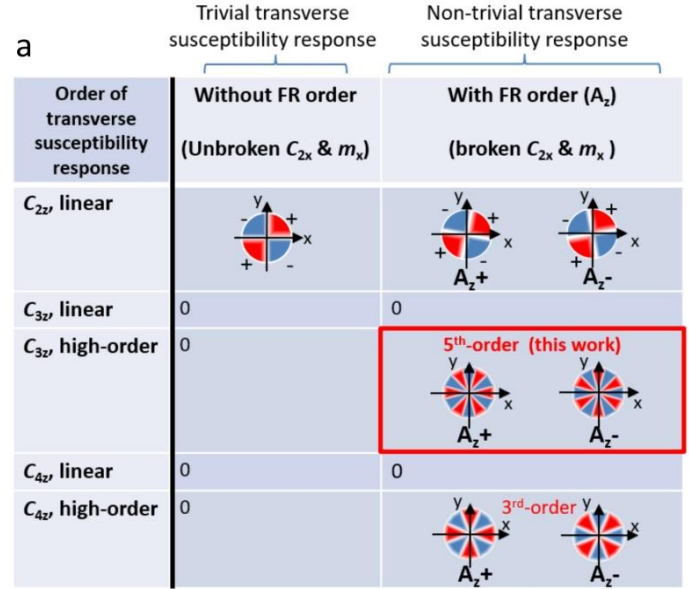

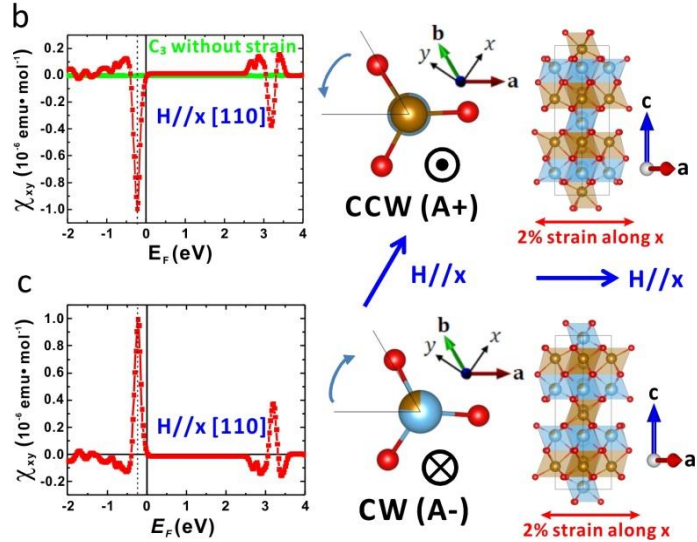

**Figure 4: Transverse susceptibility response of $C_{2z}$, $C_{3z}$, and $C_{4z}$ systems with and without FR order and first-principle calculations of ATS for $C_{3z}$.** (a) Corresponding clockwise (red regions) and counter-clockwise (blue regions) transverse response with respect to the driven field when applied within $xy$ plane. Boundaries with white color are axes with no transverse susceptibility. These "transverse response wheels" should have unbroken $C_{2x}$ and $m_x$ without FR order and broken $C_{2x}$ and $m_x$ with FR order. The sum of FR($A_z$+) and FR($A_z$-) domains will match the case without FR order. The high-order transverse electric/magnetic response is a unique characteristic of $C_{3z}$ and $C_{4z}$ systems with FR order. (b) Calculated linear transverse magnetic susceptibility $\chi_{xy}$ of CCW(A+) FR domain in FeTiO$_3$ with a 2% strain (red curve) along [110] direction parallel to the magnetic field and without strain (green curve). The linear transverse susceptibility $\chi_{xy}$ with strains along the field mimics the high-odd-order anomalous transverse susceptibility. The most pronounced signals lie around the slight hole-doped side, while the linear $\chi_{xy}$ calculation without strain has a consistent vanishing signal. (c) Calculated linear

transverse magnetic susceptibility $\chi_{xy}$ of opposite CW(A-) FR domain in FeTiO$_3$ with a 2% strain (red curve) along [110] direction parallel to the magnetic field, showing an expected sign reversal of the anomalous transverse susceptibility.

# Supplementary Information

Title

# Electrotoroidicity: New Paradigm for Transverse Electromagnetic Responses

Kai Du[1], Daegeun Jo[2], Xianghan Xu[1], Fei-Ting Huang[1], Kefeng Wang[1], David Vanderbilt[1], Hyun-Woo Lee[2], and Sang-Wook Cheong[1]*

1. Department of Physics and Astronomy, Rutgers University, Piscataway, New Jersey 08854, USA
2. Department of Physics, Pohang University of Science and Technology, Pohang, Kyungbuk 37673, Republic of Korea

* To whom the correspondence should be addressed. (E-mail: sangc@physics.rutgers.edu)

**Supplementary Information Note 1:**
**Transverse electromagnetic responses belonging to "magnetotoroidicity"**
According to our definition, the universal invariance of the orthogonal configuration among the electric dipole (***P***, or electric field ***E***), magnetic dipole (***M***, or magnetic field ***H***), and velocity vector ***k*** (or magnetic toroidal moment $T_o$[14]), termed "magnetotoroidicity", can explain all existing transverse electromagnetic responses so far. For instance, the magnetotoroidic invariance of [1] "***k*** cross ***M*** goes like ***P***" explains the (anomalous) Hall effect[1,2] and Faraday rotations in ferromagnets[3] when time-reversal (*T*) symmetry is broken. Specifically, when a longitudinal electric current ($J_x$) with the velocity ***k*** is applied to a conductor, the perpendicular external ***H*** or spontaneous ***M*** in the sample itself can break *T* symmetry and enable a transverse Hall voltage $E_y$ (Fig. 1a). Similarly, the transverse light polarization induced by Faraday rotations in ferromagnets can be considered as the anomalous Hall effect at the optical frequency[46]. [2] "***k*** cross ***P*** goes like ***M***" suggests a charge flow with the velocity ***k*** perpendicular to the polarization ***P*** of the polar system can induce a transverse magnetization[47,48] when spatial-inversion (*I*) symmetry is broken. In addition, this current-induced off-diagonal magnetization can also explain even-order nonlinear anomalous Hall effect in time-reversal invariant polar materials[4,5]. [3] "***P*** cross ***M*** goes like ***k***" indicates that systems with magnetic toroidal order can have non-reciprocal directional dichroism even for unpolarized light[7] when both *I* and *T* symmetry are broken. Finally, since magnetic toroidal order $T_o$ has the same symmetry with velocity ***k***, the first two cases ("***k*** cross ***M***" and "***k*** cross ***P***") can also clearly explain the off-diagonal linear magnetoelectricity in magnets with magnetic toroidal moments[6] when both *I* and *T* symmetries are broken. Therefore, all these transverse electromagnetic effects manifest the same magnetotoroidicity and belong to the same type of transverse responses under conditions of broken *I* or *T* symmetries (or both).

**Supplementary Information Note 2:**
**Difference between trivial and non-trivial transverse response**
We would like to elaborate the distinct difference between the trivial birefringence-type linear off-diagonal response and non-trivial "Hall-like" response such as the Hall effect and ATS in this work. In the linear regime, one can have simple and trivial off-diagonal conductivity or magnetic/electric susceptibility in in-plane anisotropic materials with two-fold rotational symmetry along *z* axis ($C_{2z}$) even without the presence of perpendicular external fields or corresponding spontaneous orders, similar to the linear birefringence effect. In this case, there exist two special orthogonal principal axes within the *xy* plane along which the applied field and the induced response can be always parallel to each other. If the field is not applied along those two principal axes, such a trivial linear off-diagonal response occurs naturally, pointing to the axis with a larger response. Apparently, this trivial linear off-diagonal response is independent of any external perpendicular fields or corresponding spontaneous orders. On the other hand, we refer to our ATS one that can be turned on by FR orders (Fig. 1b) as a non-trivial "Hall-like" off-diagonal response, in analogy to anomalous Hall effect enabled by perpendicular magnetization. Higher-order responses such as the nonlinear Hall effect and nonlinear ATS discussed in this work can go beyond the principal axes situation of the linear response since susceptibility tensors for higher-order responses cannot be fully characterized by principal axes.

In addition, we show in the next section that the $C_{3z}$ symmetry of $Fe_{1.23}Ti_{0.77}O_3$ enforces isotropic linear susceptibility ($\chi_{xx} = \chi_{yy}$), where such trivial linear off-diagonal response is forbidden. Therefore, we focus on the magnetic version of non-trivial ATS that is closely tied to the electrotoroidal moment *A* of FR orders (Fig. 1b) in this work.

**Supplementary Information Note 3:**
**Symmetry analysis of high-order anomalous transverse susceptibility (ATS)**
The $C_{3z}$ symmetry of $Fe_{1.23}Ti_{0.77}O_3$ posts additional constraints and forbids the first-order linear effect. To prove it, we take the rotation axis of the $C_3$ symmetry as the *z* axis and assume both **H** and **M** lie in the *xy* plane. In the linear response regime, the relation between **H** and **M** may be described by the second-rank magnetic susceptibility tensor χ with elements $\chi_{ij}$,

$$\begin{pmatrix} M_x \\ M_y \end{pmatrix} = \begin{pmatrix} \chi_{xx} & \chi_{xy} \\ \chi_{yx} & \chi_{yy} \end{pmatrix} \begin{pmatrix} H_x \\ H_y \end{pmatrix}. \tag{1}$$

When rotating **H** by angle $2\pi/3$ around the *z* axis, **M** should be rotated by the same angle due to the $C_{3z}$ symmetry. This leads to the constraint

$$U^\dagger \chi U = \chi, \qquad (2)$$

where $U$ is the rotation matrix by angle $2\pi/3$ around the $z$ axis,

$$U = \begin{pmatrix} \cos\frac{2\pi}{3} & -\sin\frac{2\pi}{3} \\ \sin\frac{2\pi}{3} & \cos\frac{2\pi}{3} \end{pmatrix}. \qquad (3)$$

From simple algebra, one finds from Eq. (2) the following two geometric constraints,

$$\chi_{xx} = \chi_{yy}, \qquad (4a)$$
$$\chi_{xy} = -\chi_{yx}. \qquad (4b)$$

However, the geometric constraint, Eq. (4b), is incompatible with the thermodynamic relation,

$$M_i = -\frac{\partial f}{\partial H_i}, \qquad (5)$$

where $f$ is the free energy density. In the linear response regime, Eq. (5) implies

$$\chi_{xy} = \left(\frac{\partial M_x}{\partial H_y}\right)_{H=0} = -\left(\frac{\partial^2 f}{\partial H_y \partial H_x}\right)_{H=0} = -\left(\frac{\partial^2 f}{\partial H_x \partial H_y}\right)_{H=0} = \left(\frac{\partial M_y}{\partial H_x}\right)_{H=0} = \chi_{yx}. \qquad (6)$$

Thus, from Eqs. (4b) and (6), one finds $\chi_{xy} = \chi_{yx} = 0$. That is, the off-diagonal susceptibility is forbidden in the linear response regime.

To examine the possibility that the magnetization response $M$ and the driven magnetic field $H$ may be noncollinear in the nonlinear response regime, we expand the free energy $f$ into a Taylor series of $H_x$ and $H_y$,

$$f(H_x, H_y) = f^{(0)} + \sum_{i=x,y} f_i^{(1)} H_i + \sum_{i,j=x,y} f_{ij}^{(2)} H_i H_j + \sum_{i,j,k=x,y} f_{ijk}^{(3)} H_i H_j H_k + \cdots, \qquad (7)$$

where $f_{ij}^{(2)}$ determine the magnetic susceptibility in the linear response regime whereas $f_{ijk}^{(3)}$, $f_{ijkl}^{(4)}$, and higher-order Taylor expansion coefficients determine the magnetic susceptibility in the nonlinear regime. For instance, nonvanishing $f_{yxx}^{(3)}$ implies that $H$ along the $x$ direction induces $M$ whose $y$-component is proportional to $H_x^2$. The $C_{3z}$ and time-reversal symmetries impose severe constraints on the Taylor expansion coefficients. The time-reversal symmetry requires $f_i^{(1)}$, $f_{ijk}^{(3)}$, and all odd-order expansion coefficients to vanish. To find out the implication of the $C_{3z}$ symmetry, we switch independent variables from $H_x$ and $H_y$ to $H_\pm = H_x \pm iH_y$,

$$f = f^{(0)} + \sum_{i,j=\pm} g^{(2)}_{ij} H_i H_j + \sum_{i,j,k,l=\pm} g^{(4)}_{ijkl} H_i H_j H_k H_l + \cdots, \tag{8}$$

where odd-order terms are absent due to time-reversal symmetry. Under the $C_{3z}$ rotation, $H_\pm$ transforms as

$$H_\pm \to e^{\pm \frac{i2\pi}{3}} H_\pm. \tag{9}$$

Considering that all terms in Eq. (8) can be expressed as $H_+^m H_-^n$ with $m+n$ being a non-negative even integer, we obtain the general transformation rule under the $C_{3z}$ rotation,

$$H_+^m H_-^n \to e^{i(2\pi/3)(m-n)} H_+^m H_-^n. \tag{10}$$

Thus, for $H_+^m H_-^n$ to remain invariant under the $C_{3z}$ rotation, the factor $e^{i(2\pi/3)(m-n)}$ should be one, and $m-n$ should be integer multiples of three. Together with the constraint from the time-reversal symmetry that $m+n$ should be a non-negative even integer, one finds that $m-n$ should be integer multiples of six. For $m=n$, the resulting term, $H_+^m H_-^n = (H_x^2 + H_y^2)^n$, is isotropic with respect to the **H** direction change and does not contribute to the off-diagonal susceptibility. For $m \neq n$, on the other hand, the resulting term may give rise to the off-diagonal susceptibility. Considering that $m-n$ should be integer multiples of six, the lowest-order terms of $f$ that contribute to the off-diagonal susceptibility appear in the sixth-order terms of the Taylor series expansion. Up to this order, $f$ is given by

$$f = f_{\text{iso}} + f_{\text{ani}} + O(H_+, H_-)^8, \tag{11}$$

where

$$f_{\text{iso}} = f^{(0)} + g^{(2)}_{\text{iso}}(H_x^2 + H_y^2) + g^{(4)}_{\text{iso}}(H_x^2 + H_y^2)^2 + g^{(6)}_{\text{iso}}(H_x^2 + H_y^2)^3, \tag{12a}$$

$$f_{\text{ani}} = g^{(6)}_+ H_+^6 + g^{(6)}_- H_-^6. \tag{12b}$$

Here, $f_{iso}$ is isotropic with respect to the **H** direction change and does not contribute to the off-diagonal susceptibility. In contrast, $f_{ani}$ varies with the **H** direction (anisotropy) and does contribute to the off-diagonal susceptibility. To verify the latter point, we first utilize the relation $g_-^{(6)} = \left(g_+^{(6)}\right)^*$ due to the constraint that $f_{ani}$ is real, and obtain

$$f_{ani} = 2\text{Re}\left[g_+^{(6)} H_+^6\right]. \tag{13}$$

We then express $f_{ani}$ in terms of $H_x$ and $H_y$,

$$\begin{aligned}f_{ani} = &\ 2\text{Re}\left[g_+^{(6)}\right]\left(H_x^6 - 15H_x^4 H_y^2 + 15H_x^2 H_y^4 - H_y^6\right) \\ &- 2\text{Im}\left[g_+^{(6)}\right]\left(6H_x^5 H_y - 20H_x^3 H_y^3 + 6H_x H_y^5\right).\end{aligned} \tag{14}$$

From Eqs. (12a) and (14), one finds that $M_y = -\partial f/\partial H_y$ is given by $12\text{Im}\left[g_+^{(6)}\right] H_x^5$ when **H** is along the *x*-direction. This verifies that the off-diagonal susceptibility is possible in systems with the $C_{3z}$ symmetry, and the leading order contribution to the off-diagonal **M** response arises in the 5th order, that is,

$$M_y = \chi_{yxxxxx} H_x^5 \quad \text{with} \quad \chi_{yxxxxx} = 12\text{Im}\left[g_+^{(6)}\right]. \tag{15}$$

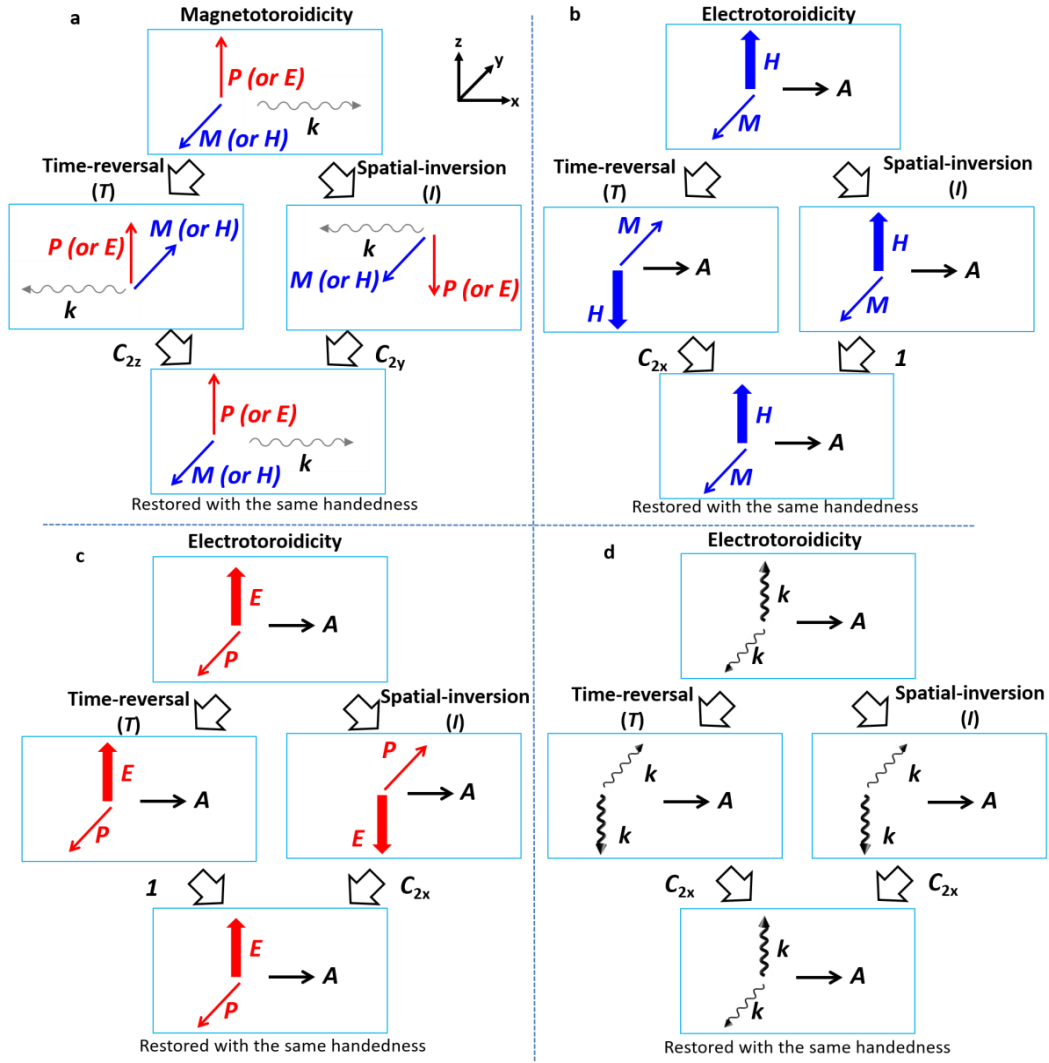

**Fig. S1.**

**Symmetry-invariant magnetotoroidicity and electrotoroidicity.** (**a**) The orthogonal configuration among the electric dipole (**P**, or electric field **E**), magnetic dipole (**M**, or magnetic field **H**), and velocity vector **k** (or magnetic toroidal moment $T_o$) is invariant under time-reversal and spatial-inversion operations, which can be restored by simple two-fold rotations ($C_{2z}$ or $C_{2y}$). The invariance of this configuration is termed "magnetotoroidicity". (**b**) The orthogonal configuration among the magnetic field **H**, magnetic dipole **M**, and electrotoroidal moment **A** is invariant under time-reversal and spatial-inversion operations, which can be restored by the simple two-fold rotation ($C_{2x}$). (**c**) The orthogonal configuration among the electric field **E**, electric dipole **P**, and electrotoroidal moment **A** is invariant under time-reversal and spatial-inversion operations, which can be restored by the simple two-fold rotation ($C_{2x}$). (**d**) The orthogonal configuration among the applied velocity vector **k** (thicker curved arrow), induced transverse velocity vector **k** (thinner curved arrow), and electrotoroidal moment **A** is invariant under time-reversal and spatial-inversion operations, which can be always restored by the simple two-fold rotation ($C_{2x}$). The invariance of configurations in (b-d) is termed "electrotoroidicity", which corresponds to a new type of transverse electromagnetic responses.

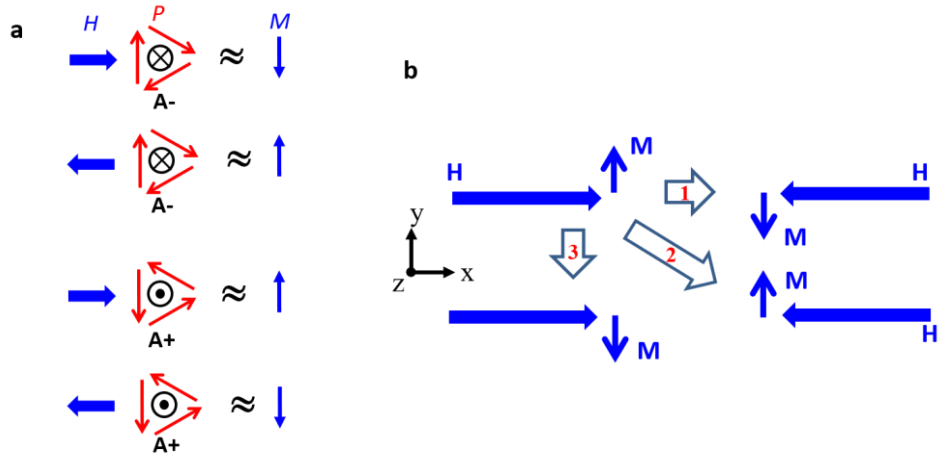

**Fig. S2.**
**Odd-order anomalous transverse magnetic susceptibility and its symmetry requirements.** (a) Odd-order anomalous transverse magnetic susceptibility of ferro-rotational order, where the off-diagonal response changes its sign upon switching the ferro-rotational order or the magnetic field. (b) To have odd-order off-diagonal $\chi_{xy}$, one needs to preserve the symmetry operations for the identity and route path 1 and break the symmetry operations for the route path 2 and route path 3. That is unbroken $\{1, C_{2z}, m_z, I, T\}$ and broken $\{C_{2x}, C_{2y}, m_x, m_y\}$, which means broken $\{1, C_{2z}, m_z, I, T\} \otimes \{C_{2x}, C_{2y}, m_x, m_y\}$=broken $\{C_{2x}, C_{2y}, m_x, m_y, C_{2x}\otimes T, C_{2y}\otimes T, m_x\otimes T, m_y\otimes T\}$, where $m$ is the mirror reflection perpendicular to a certain axis, $C_2$ is the two-fold rotation along a certain axis, $I$ is the inversion, $T$ is the time reversal, and $\otimes$ stands for the symmetry operation on the left plus the one on the right.

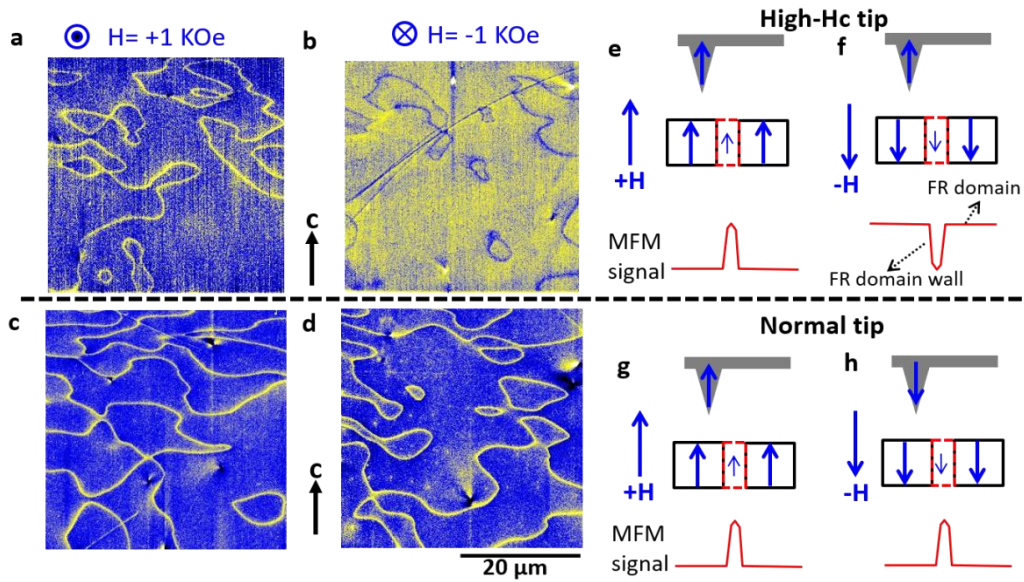

**Fig. S3.**

**Confirming reduced diagonal susceptibility at FR domain walls with different MFM tips.** **(a-b)** MFM images taken by non-switchable high-coercivity tip under 1 kOe opposite out-of-plane fields and **(c-d)** their corresponding schematic signals showing domain wall signal reversals. **(e-f)** MFM images taken by switchable normal MFM tip under 1 kOe opposite out-of-plane fields and **(g-h)** their corresponding schematic signals showing the same domain wall signal.

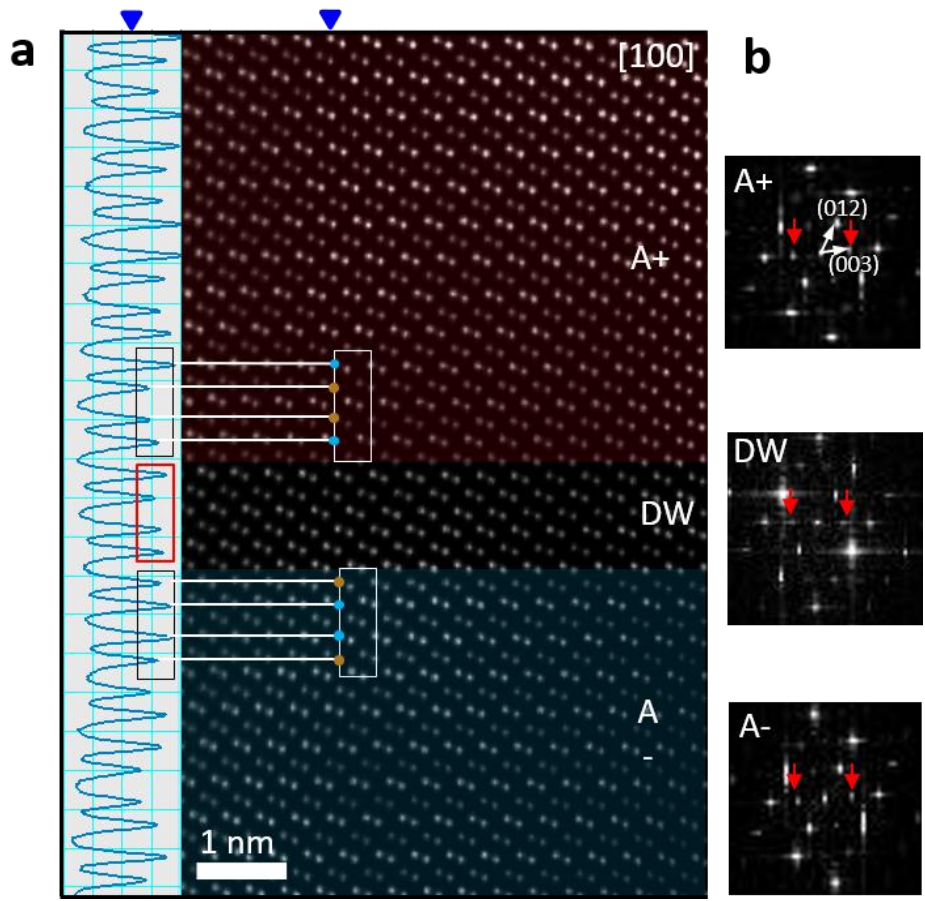

**Fig. S4.**

**Atomic imaging of ferro-rotational domains and the domain wall.** (**a**) High-angle annular dark-field STEM image taken along [100], accompanied by the extracted intensity profile along the single cation column denoted by a blue arrowhead. The arrangement of "bright-dark-dark-bright" layers stacked alternately along the crystallographic *c*-axis illustrates the ordered ilmenite structure. The positions showing higher and lower intensity correspond to Fe and Fe-Ti columns, respectively. The upper region, exhibiting a systematic variation of "bright-dark-dark-bright" is attributed to an A+ domain, while the lower region, displaying "dark-bright-bright-dark" signifies the A- domain. The red rectangle highlights the intensity profile at the domain wall where a reversal of the periodicity in intensity patterns occurs. The domain wall is about one-unit cell in width. (**b**) The FFT diffractograms obtained from the selected areas of top (A+), domain wall (DW), and bottom (A-) regions show the persistence of (003) peaks (red arrows). Though the intensity of (003) peaks is significantly reduced, we confirm the ilmenite structure at the DW instead of the formation of corundum structure with completely random distribution of cations. However, a smaller intensity variation at the DW and a relatively darker DWs in comparison to domains imply a slight Fe deficiency at DWs in our $Fe_{1.23}Ti_{0.77}O_3$ single crystals.

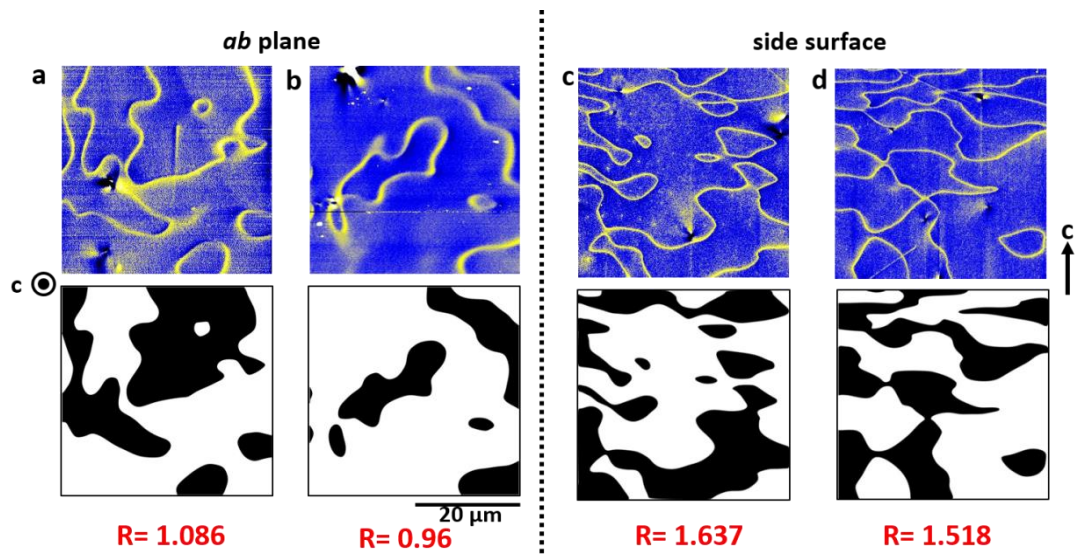

**Fig. S5.**

**FR domain shape anisotropy. (a-b)** MFM images of isotropic FR domains on the *ab* plane and **(c-d)** MFM images of laterally elongated FR domains on the side surface. One can convert MFM images into black and white domain images and automatically calculate the ratio (R) between the averaged lateral domain size and the averaged vertical domain size following the previously reported method[19]. The ratio R close to 1 on the *ab* plane indicates an isotropic FR domain shape and the ratio R larger than 1 on the side surface reflects a laterally elongated FR domain shape.

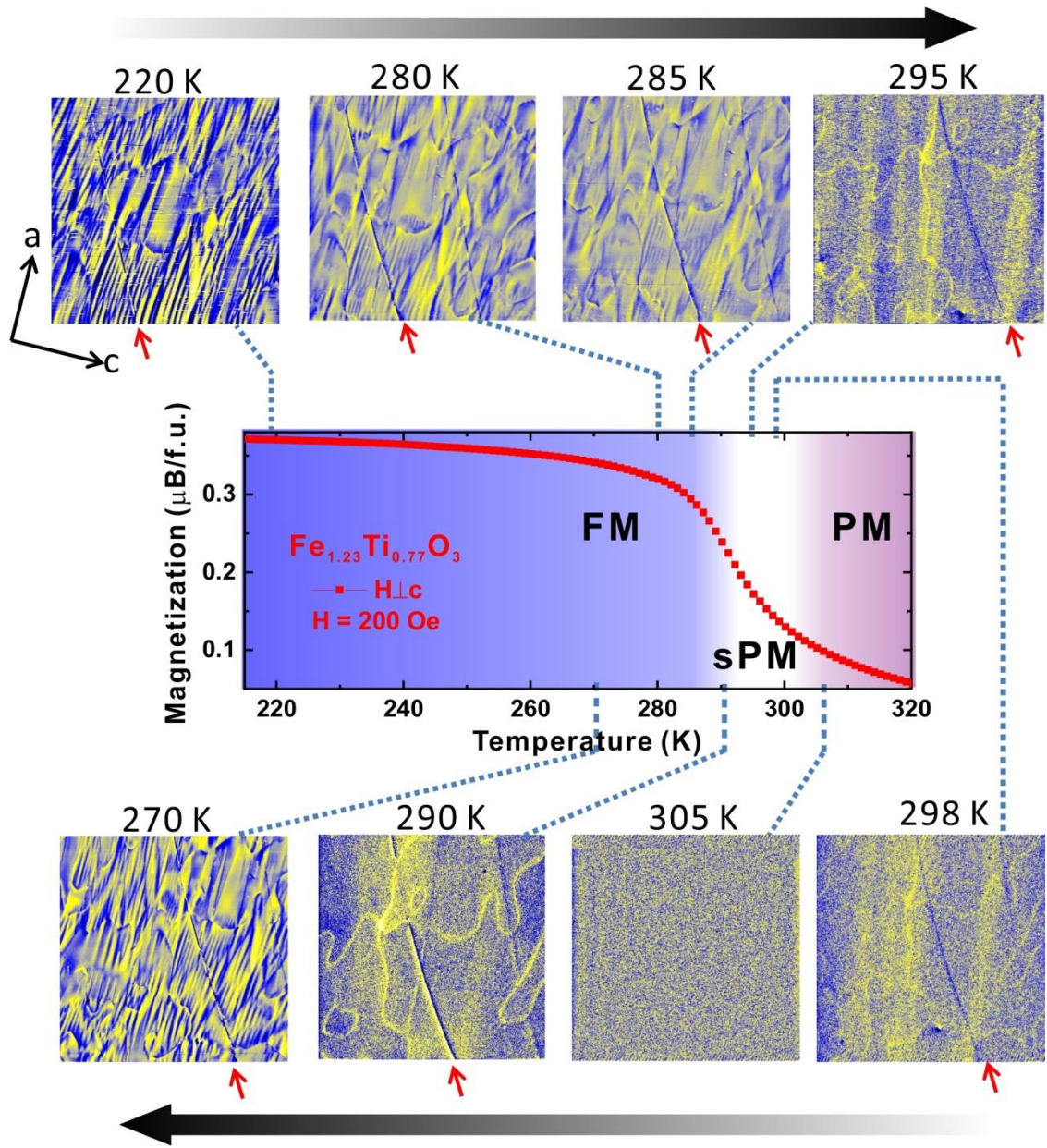

**Fig. S6.**

**Temperature-dependent MFM images down to low temperatures.** Abrupt terminations of ferrimagnetic domains at FR domain walls are observed at low temperatures, which demonstrates a strong coupling between the FR and ferrimagnetic order and possible exotic domain wall magnetism in magnetic FR materials.

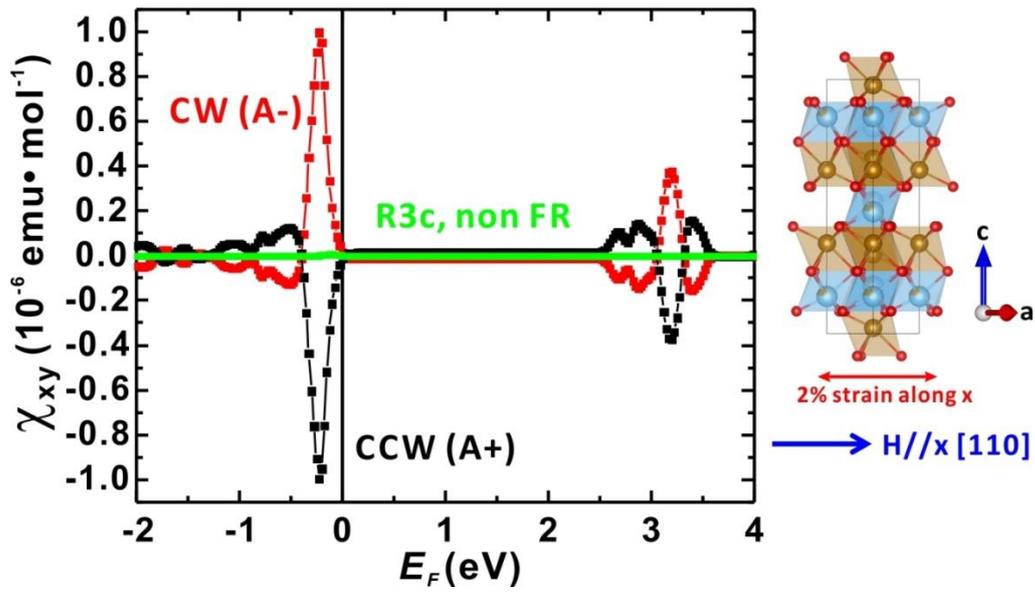

**Fig. S7.**

**First-principle calculations of high-odd-order anomalous transverse magnetic susceptibility with and without FR order.** Calculated linear transverse magnetic susceptibility $\chi_{xy}$ of CCW(A+) FR domain (black curve) and CW(A-) FR domain (red curve), as well as the non-FR $R$3c structure (green curve) in FeTiO$_3$ with a 2% strain along [110] direction parallel to the magnetic field. The vanishing $\chi_{xy}$ of non-FR $R$3c structure exemplifies the critical role of FR order for anomalous transverse susceptibility.

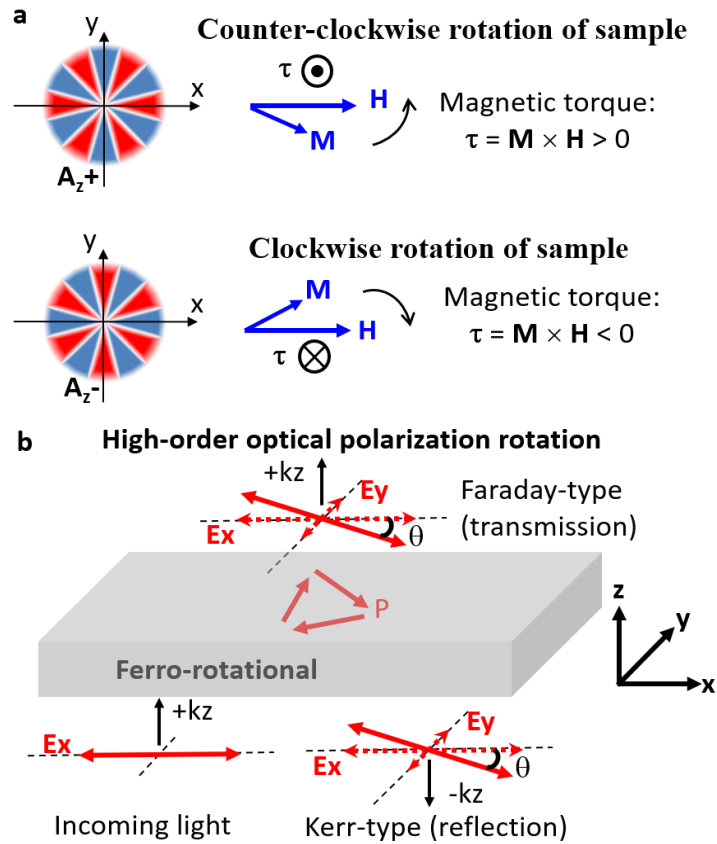

**Fig. S8.**

**New functionalities of anomalous transverse susceptibility in ferro-rotational materials.** **(a)** Magnetic torques excerted to single-domain ferro-rotational samples in the presence of magnetic fields, which can be ultilized to rotate/manipulate ferro-rotational particles or construct continuous rotating ferro-rotational motors in AC magnetic fields. **(b)** Possible high-order optical rotations of linearly-polarized light of ferro-rotational systems in both transmission (Faraday-type) and reflection (Kerr-type) geometries, in contrast to the natural optical rotation in chiral materials that only occur in the transmission geometry.